\documentclass[10pt,reqno]{amsart}
\usepackage[centertags]{amsmath}
\usepackage{multirow}
\usepackage{rotating}
\usepackage{graphicx}
\usepackage{epsfig}

\theoremstyle{plain} 

\theoremstyle{definition}

\theoremstyle{remark}

\numberwithin{equation}{section}

\DeclareMathOperator{\Max}{Max}

\DeclareMathSymbol{\R}{\mathalpha}{AMSb}{"52}
\DeclareMathSymbol{\C}{\mathalpha}{AMSb}{"43}

\newcommand{\ra}{\rightarrow}

\begin{document}

\title[Implicit methods for barrier options]{
High-order accurate implicit methods for the pricing of barrier
options}
\author[Ndogmo, J.C.]{Ndogmo, J.C.}

\address{Department of Physics\\
Univ. of Western cape\\
Private Bag X17\\
Bellville 7535\\
South Africa.}
\email{jndogmo@uwc.ac.za}

\author[Ntwiga, D.B.]{Ntwiga, D.B.}

\address{NBN\\
Univ. of Western cape\\
Private Bag X17\\
Bellville 7535\\
South Africa.}
\email{ntwiga@nbn.ac.za}


\begin{abstract}
This paper deals with a high-order accurate implicit
finite-difference approach to the pricing of barrier options. In
this way various types of barrier options are priced, including
barrier options paying rebates, and options on
dividend-paying-stocks. Moreover, the barriers may be monitored
either continuously or discretely. In addition to the high-order
accuracy of the scheme, and the stretching effect of the coordinate
transformation, the main feature of this approach lies on a
probability-based optimal determination of boundary conditions. This
leads to much faster and accurate results when compared with similar
pricing approaches. The strength of the present scheme is
particularly demonstrated in the valuation of discretely monitored
barrier options where it yields values closest to those obtained
from the only semi-analytical valuation method available.
\end{abstract}

%
\maketitle

\section{Introduction}
\label{s:intro} Barrier options are a type of path-dependent options
whose values depend on the specific path followed by the underlying
asset during the option's life, and w.r.t. some specified asset
values, usually referred to as barriers. These options are exotic
derivatives traded in over-the-counter markets and they can
therefore be tailored to specific customer needs. Moreover, the
additional constraints imposed on them by the barrier makes them a
much cheaper and attractive product then the standard options.
Barriers can also be added to any existing type of standard or
exotic option.  The barrier option market has thus been expanding
rapidly and it has been estimated \cite{hsu} that it has doubled
every year since 1992. In parallel with this development, a large
number of newly designed barrier options have been trading very
actively in financial markets  \cite{car, zvan}.
\par

Analytical formulas exist for most of the standard barrier options.
Rubinstein and Reiner \cite{rub}, Rich \cite{rich}, and Kunitomo and
Ikeda \cite{kuni} derived analytical formulas for a variety of
standard knock-in and knock-out European options with full barriers.
Heynen and Kat \cite{heynen} obtained similar formulas for some
special types of barrier options, namely the partial barrier
options, and for so-called outside barrier options where it is
another variable different from the underlying asset which
determines whether the option knocks in or out. Closed-form
solutions for double barrier options, some of which are
time-dependent, have been obtained \cite{kuni, geman, pelsser}. The
extension of these formulas to the more complex case of
American-style barrier options has been undertaken by Broadie and
Detemple \cite{broadie}, Gao {\em et al.} \cite{gao}, as well as
Haug \cite{haug2}. However, all the formulas obtained in such
extensions are only closed-form approximations.
\par

Despite all these efforts to derive analytical pricing formulas for
barrier options, there is still no analytical formulas available for
large classes of both European- and America-style barrier options
\cite{haug2, kuni}.  Although Merton \cite{merton} proposed the
closed-form solution for the continuously monitored down-and-out
barier option in 1973, it's only in the $1990$'s that valuation
formulas were obtained for other variants of this European-style
option \cite{rub, rich, heynen, kuni}. This is simply because the
valuation of financial options has led to mathematical models which
are most often challenging to solve. For instance, for most of the
complex options, there are no analytical formulas available, and
almost all formulas have been obtained under the assumptions that,
amongst others, the underlying asset price has a lognormal
distribution with constant drift and volatility parameters, and that
there are no transaction costs  \cite{rub, haug2}. However, as
empirical evidence suggest the contrary, new models with stochastic
or time-dependent volatilities, or including transaction costs have
been designed and implemented. Nevertheless, these attempts to
improve on the assumptions underlying the derivation of analytical
formulas have been made mostly for standard American options
\cite{heston, during, ikonen}. \par

For the valuation of options, and exotic options in particular,
numerical methods remain a tool of choice. This is first justified
by the fact that the number of exotic options that can be designed
is limitless, and as new products enter the market one of the most
practical way, if not the only available means to price them, are
quite often the numerical methods. They are also a convenient
benchmark for testing the validity of analytical formulas,
especially as they become available. There are even instances where
they are faster than analytical methods, even those analytical
methods involving fast converging infinite sums, when the required
accuracy isn't too high.
\par

As a numerical method for option valuation, the lattice methods have
been used extensively, and for barrier options it appears however
that they are essentially useless without the proper positioning of
the barrier which must line-up with the tree nodes. Despite all the
techniques that have been developed to improve on lattice methods,
they are still quite computer-intensive \cite{haug2}. Monte Carlo
Methods have been considered as inflexible and unreliable for option
pricing, until the last decade where they yielded more promising
results based on innovative techniques which are now a topic of
current research. Barraquand and Martineau \cite{barraq} amongst
others, obtained in this context some interesting results for
multi-asset American options. The most commonly used numerical
method in option pricing these days appears to be the finite
difference methods. They have an acceptable computational cost when
an appropriate implicit method is used (see \cite{zvan, mayo,
ikonen}). They can also be extended to cater for American-style
options, by formulating them as linear complementary problems. \par

In this paper we consider an implicit finite difference approach
   to the valuation of barrier options. These barrier options may
   have various features including double barriers,
   dividend-paying-stocks, rebate payments, or some
   combination of these. Barriers may be monitored either continuously or
   discretely. We use a well-known $\theta$-method which is
   fourth-order accurate in space and second-order accurate in time,
   by relating the parameter $\theta$ to the mesh sizes. In addition
   to the high-order accurate scheme and the stretching effect of
   the coordinate transformation used, the strength of this approach
   lies on a probability-based optimal determination of the boundary
    conditions, along
   which the option values are known exactly. In this way, with a
   reasonable accuracy and for the most practical considerations,
   we are generally able to use fewer than
   20 asset prices and 20 time steps per year to value most of the barrier
   options, especially when the barrier is continuously applied.
   However, the efficiency of our approach is much clearly
   demonstrated in the case of discretely monitored barrier options,
when we compare the results we obtained with that of several other
authors. As usual, due to some parity considerations, we shall focus
our attention only on knock-out call options.\par

   This paper is organized as follows. In Section \ref{s:model} we discuss the
mathematical model for valuing barrier options. Section \ref{s:fds}
is devoted to the discretization of the modeling differential
equation, while Section \ref{s:nsol} discusses applications of the
resulting implicit scheme to the various types of barriers options,
and Section \ref{s:nres} reports the corresponding numerical
results.

\section{Option pricing model}\label{s:model}
  We make the usual assumption that the underlying asset price $S$
  has a geometric Brownian motion, with drift and volatility
  parameters $\mu$ and $\sigma,$ respectively. Thus the process
  followed by $S$ can be written with the usual notation and in
  terms of the Wiener process $W$ as

\begin{equation} \label{eq:2.1} d S= \mu dt + \sigma d W. \end{equation}

The price $f=f(S,t)$ of a European option contingent on $S$ must
then satisfy the Black-Scholes differential equation

\begin{equation} \label{eq:2.2} \mathcal{L} \cdot f = 0 \end{equation}
where the partial differential operator $\mathcal{L}$ is given by
\begin{equation} \label{eq:2.3} \mathcal{L}:= \frac{\partial}{\partial t} + \mu
S \frac{\partial}{\partial S} + \frac{1}{2} S^2 \sigma^2
\frac{\partial^2}{ \partial S^2} - r, \end{equation}
and where $r$ denotes the risk-free interest rate. We shall usually
assume that the underlying asset price pays a continuous dividend
yield at a continuous rate of $q$ per year, and as we are placing
ourselves in the traditional risk-neutral world in which the market
price of risk is zero, we shall have in this case $\mu= r-q.$
\par

Thanks to the put-call parity for European options,  and also to the
similar parity relation between knock-ins and knock-outs, we shall
consider only knock-out call options. To begin with, let $B$ denote
the constant barrier value for a down-and-out barrier option. Since
the value of the option at expiration time $T$ is its payoff $\Max
(S-K,0),$ where $K$ is the strike price of the option, the initial
condition for equation \eqref{eq:2.2} is given by
\begin{equation} \label{eq:2.4} f(S, T)= \Max (S-K, 0), \qquad \text{for $S >
B.$} \end{equation}
The corresponding boundary conditions follow from the properties of
call options. In the simplest case where the constant barrier value
$B$ is continuously applied, denoting by $Rb$ the rebate received if
the barrier is ever breached, the boundary conditions are

\begin{align}
f(B,t) & = Rb, \qquad \forall t \label{eq:2.5} \\
f(S,t) & \sim S, \qquad \forall t, \qquad \text{ as $S \ra \infty
$}. \label{eq:2.6}
\end{align}
It should be noted that boundary condition \eqref{eq:2.6} which is
usually used for the valuation of call options is not suitable for
an optimal algorithm, in particular because the value of $S$ in that
condition is generally exceedingly large, and the resulting equality
is only an approximation, and all these tend to slow down the
valuation algorithm. We shall therefore always strive in this paper
to find optimal boundaries of the solution domain along which the
option values are known exactly, and which also yield an optimal
truncation of the solution domain. \par

 The Black-Scholes equation \eqref{eq:2.2} can be analyzed directly for the valuation
of barrier options, as for example in  \cite{boyle98} and
\cite{zvan}. However, there is a numerically more effective
coordinate transformation that we shall adopt here, and which is
obtained by first setting
\begin{align}
x & = \ln (S/K), \label{eq:2.7} \\
\tau & = \frac{\sigma^2}{2} (T-t). \label{eq:2.8}
\end{align}
Since for all practical considerations the underlying price range
generally lies in a subinterval of $\left(  K/3, 3K \right),$ it
appears that \eqref{eq:2.7} restricts the underlying asset price in
the $x$-coordinate to range most often in a small interval such as
$\left( -1.099, 1.099\right).$ On the other hand, \eqref{eq:2.8} not
only transforms  \eqref{eq:2.2} into a forward-time problem, but it
also has the advantage of shrinking the time range. Under the change
of coordinates $\eqref{eq:2.7}$ and \eqref{eq:2.8}, equation
\eqref{eq:2.2} is transformed into
$$
\frac{\partial f}{ \partial \tau } = \frac{\partial^2 f}{\partial
x^2} + (\nu -1)\frac{\partial f}{\partial x} - \nu_1 f
$$
where $\nu= 2 \mu / \sigma^2,$ and  $ \nu_1=  2 r / \sigma^2.$ So,
$\nu = \nu_1-\nu_2,$ where $\nu_2= 2 q/ \sigma^2$. Here $\nu_1$ and
$\nu_2$ are the only two non-dimensional parameters of the
Black-Scholes equation for dividend-paying-assets. To discard the
terms in $f$ and $\partial f / \partial x$ in this last differential
equation, we now make a change of the dependent variable by setting

\begin{equation} \label{eq:2.9} f (S,t)= K e^{\alpha x + \gamma \tau} u (x,
\tau), \end{equation}
where $\alpha = -\frac{1}{2} (\nu -1)$ and $\gamma = -\frac{1}{4}
(\nu +1)^2 - \nu_2.$ This last change of variables reduces the
original equation \eqref{eq:2.2} to the diffusion equation

\begin{equation} \label{eq:2.10} \frac{\partial u}{\partial \tau} =
\frac{\partial^2 u}{ \partial x^2}. \end{equation}

Letting $x_b= \ln(B/K),$ the initial and boundary conditions given
earlier are also transformed into the following expressions in the
$(x, \tau)$-coordinates.
\begin{subequations}
\begin{align}
u(x,0) & = \Max \left( e^{(1/2)(\nu +1)x} - e^{(1/2)(\nu-1)x}, 0
\right),
\qquad x > x_b \label{eq:2.11a} \\
u(x, \tau ) & \sim e^{(1-\alpha)x -\gamma \tau}, \qquad \text{as $x
\ra \infty$}   \label{eq:2.11b}\\
\intertext{and}
 u(x_b, \tau) &= \frac{Rb}{K} \; e^{- ( \alpha x_b + \gamma \tau)}.
 \label{eq:2.11}
\end{align}
\end{subequations}
It is worth nothing at this point that the change of variable
\eqref{eq:2.9} is also very convenient for an efficient numerical
algorithm. Indeed, it gives the final solution as a scalar multiple
of the numerically computed factor $u (x, \tau),$ where the scalar
$K e^{\alpha x + \gamma \tau}$ is known exactly and so, does not
involve any error. In this way any error expansion due to
transformation \eqref{eq:2.9} will tend to be minimized.


\section{Finite difference scheme}
\label{s:fds}

Denote by $h=\Delta x$ and $k= \Delta \tau$ the mesh sizes in the
space and the time directions, respectively. Denote also by $U_j^n$
the discrete approximation of $u(x, \tau)$ at the grid node $(x_j,
t_n), $  where $x_j= j \Delta x,$  and $t_n= n \Delta t.$  Let
$\delta_x^2$  be the difference operator given by $\delta_x^2 U_j^n
= U_{j-1}^n - 2 U_j^n + U_{j+1}^n.$ For the discretization of the
reduced equation \eqref{eq:2.10} we choose the two-time level,
three-space-point scheme
\begin{equation} \label{eq:3.1} \frac{U_j^{n+1} - U_j^n}{\Delta t} =
\frac{1}{(\Delta x)^2} \left( \theta \,\delta_x^2 U_j^{n+1} + (1-
\theta)\, \delta_x^2 U_j^n\right) \end{equation}
also called weighted average approximation or $\theta$-method. The
weight $\theta$ in this expression is assumed to satisfy the
condition $0 \leq \theta \leq 1,$ in order to avoid negative
weights. The values $\theta= 0$ and $\theta=1$ give the explicit
scheme and the fully implicit scheme, respectively. If we let $\beta
= \Delta t / (\Delta x)^2,$ i.e. $\beta = k /h^2,$ then for $0 \leq
\theta < \frac{1}{2},$  the scheme \eqref{eq:3.1} is stable if and
only if $\beta \leq     1/ 2(1- 2 \theta).$ For $\frac{1}{2} \leq
\theta \leq 1,$ it is stable for all values of $\beta.$ For this
scheme, the truncation error, i.e. the amount by which the exact
solution of the differential equation does not satisfy the discrete
approximation \eqref{eq:3.1}, is generally of first order accuracy
in $\Delta t,$ while the popular Crank-Nicolson scheme which
corresponds to $\theta = 1/2$ is second-order accurate in both
$\Delta t$ and $\Delta x.$
\par

For a much higher accuracy of the finite difference scheme ({\sc
fds}) \eqref{eq:3.1}, we shall relate in this paper the choice of
the parameter $\theta$ to the values of $\Delta x$ and $\Delta t$ of
the grid line spacings. More precisely, we shall assume that
\begin{equation}  \label{eq:3.2}  \theta= \frac{1}{2} - \frac{1}{12 \beta}. \end{equation}
It is well-known  that the resulting scheme is $O \left((\Delta t)^2
+ (\Delta x)^4\right),$ i.e. second- order accurate in time and
fourth order accurate in space (see e.g. \cite{morton}). Such a
scheme has been used in \cite{mayo} for the valuation of American
options with error correction at the critical boundary, leading to
results that are comparatively more accurate than the standard ones
found in the literature. However, this specific weighted average
scheme has not yet been applied to the valuation of barrier options,
to the best of our knowledge. Zvan {\em et al} used a method called
point-distributed finite volume scheme in \cite{zvan} to directly
discretize the Black-Scholes equation, but that method is only first
order accurate in time, and the paper does not comment on the speed
(in terms of number of time steps) of the scheme for achieving a
given level of accuracy.  \par

Since $\beta$ is positive, condition \eqref{eq:3.2} restricts
$\theta$ to the interval $[0, 1/2),$ but the condition for stability
$\beta \leq 1/ (2(1- 2 \theta)$ is always satisfied in this case.
Moreover, the condition $0 \leq \theta$ implies that
\begin{equation}\label{eq:kh}(\Delta x)^2 \leq 6 \Delta t.\end{equation}
This in practical terms means that even by requesting \eqref{eq:3.2}
to hold, we can still take large time steps while maintaining
accuracy and stability.\par

If we denote by $M$ and $L$ the number of space- and time-intervals,
respectively, in the solution grid, and by $W_{h,k}$ the finite difference
operator given by
$$
W_{h,k} \cdot U_j^n = - \beta \theta U_{j-1}^n + (1+ 2 \beta \theta)
U_j^n - \beta \theta U_{j+1}^n,
$$
then an expansion of the discretized equation \eqref{eq:3.1} leads
to
 a linear algebraic system of equations of the form

\begin{equation} \label{eq:3.3}
 W_{h,k} \cdot U_j^{n+1} = W_{h,k} \cdot U_j^n,
\qquad \text{for $n=0, \dots, L-1$ and $j=1, \dots, M-1.$}
\end{equation}

The system \eqref{eq:3.3} can be written in terms of tridiagonal
matrices and we shall solve it with  the Thomas algorithm which
includes Gaussian elimination without pivoting.
\section{Numerical solution for barrier options}
\label{s:nsol}

We discuss in this section the application of the {\sc fds}
described above to the derivation of a numerical solution to the
pricing problem for barrier options.

\subsection{Overview}
\label{s:overview}
\par
As already indicated, we shall always implement whenever possible in
this paper  an optimal truncation of the solution domain by
replacing the approximate boundary condition of the form
\eqref{eq:2.6}, or \eqref{eq:2.11b} equivalently, with a boundary
condition for which both the option value is known exactly and the
resulting solution domain is the smallest possible. Let $S_0$ be the
initial price of the underlying and $S=S_{t}$ the price of the
underlying at a time $t, ~0<t\leq T.$ In the case of a down-and-out
call option, the probability $P(S_{t}>B)$ that the barrier will not
be breached at time $t$ is given by
\begin{equation} \label{eq:4.1} P(S_{t} > B) = \Phi (a^{*}),\end{equation}
where $\Phi$ is the standard normal distribution function and
$$a^{*} = (\ln(S_{0}/B) + \mu^{*}t)/\sigma \sqrt{t}, \qquad
\text{ with } \mu^{*}=\mu - \sigma^{2}/2.$$
 Let $\delta$ be the smallest number such that $\Phi(\delta)$ has
 numerical value $1.$ The value
of $\delta$ depends on the level of accuracy required and also on
the computing system used for the calculations. An appropriate value
for $\delta$ usually lies in the interval $(3.7, 6.5),$ depending on
the level of acuracy required. For a given parameter set, the
minimum value of $S_{0}$ which guarantees that the barrier will not
be breached at time $t$ is given by the inequality $a^{*}\geq
\delta,$ or equivalently by $S_{0}\geq B e^{\delta \sigma \sqrt{t} -
\mu^{*} t}.$ Now, let $t_{p}$ be the turning point of the function
$t \mapsto  \delta \sigma \sqrt{t} - \mu^{*}t,$ and denote by
$S_{m}$ the minimum value of the current stock price $S_0$ above
which the barrier becomes worthless, and set $x_{m}=\ln (S_{m}/K).$
Then it is easy to see that

\begin{equation} \label{eq:4.2}
x_m = \begin{cases} x_b +  \delta \sigma \sqrt{T} - \mu^*T, & \text{
if
$\mu^*  \leq 0$ or $t_p \geq T$ }\\
x_b + \delta \sigma \sqrt{t_p} - \mu^* t_p , & \text{ otherwise.}
\end{cases}
\end{equation}

For any value of $S_{0}$ above $S_{m},$ or equivalently for any
value of $x_{0}$  above $x_{m},$ where $x_{0}=\ln(S_{0}/K),$ the
barrier is worthless and the option behaves exactly as a standard
option. In particular, the corresponding option value along the
boundary $S=S_{m}$ or above this boundary is given by the well-known
Black-Scholes formula for European call options.

It therefore follows from Equations \eqref{eq:2.5}, \eqref{eq:2.8}
and \eqref{eq:4.2} that the solution domain for $u(x,\tau)$ reduces
to the rectangle $[x_{b},x_{m}]\times [0,\frac{\sigma^{2}}{2} T].$
Similarly for an up-and-out barrier option, if we let $S_{m}$ denote
the maximum value of the current underlying asset price below which
the barrier becomes worthless, then $x_{m} = \ln(S_{m}/K)$ is given
by

\begin{equation} \label{eq:4.3} x_{m} = \begin{cases} x_b - \delta \sigma
\sqrt{T} -
\mu^* T, & \text{ if $\mu^* \geq 0$ or $t_p \geq T$} \\
x_b - \delta \sigma \sqrt{t_p} - \mu^* t_p, & \text{ otherwise.}
\end{cases}
\end{equation}

In this case the optimal solution domain reduces to the rectangle
$[x_{m}, x_{b}]\times[0,\frac{\sigma^{2}}{2} T ].$  This optimal
determination of the boundary conditions has several applications,
including the classification of double barrier options. A similar
determination of optimal boundary conditions can be achieved for
time-varying barriers, as opposed to the constant barriers we have
discussed here. Note that by construction we have $x_m > x_b$ in
\eqref{eq:4.2} and $x_m < x_b$ in \eqref{eq:4.3}\par

Our determination of the numerical solution for a pricing problem
will amount to calculating $u(x_{0}, \tau_{M}),$ where
$\tau_{M}=\frac{\sigma^{2}}{2} T,$ and this may require
interpolation if $x_{0}$ does not line up with a grid line. However,
for $M$ large, it is always possible to slightly enlarge the range
of the spacial variable $x$ without affecting the performance of the
algorithm in any way, just by adjusting the value of $x_{m}$ by a
fraction of the grid increment By so doing, we can always line up
$x_{0}$ with a grid point, and such adjustment of $x_{m}$ is
feasible when for instance the grid increment $h=\Delta x$ is small.
We shall therefore combine these two techniques in our calculation
of $u(x_{0},\tau_{M}).$ \par

The price of barrier options often displays some singularities near
the barrier value and it is therefore customary in a finite
difference pricing approach for these options to use adaptive grids
in an attempt to achieve the required accuracy of the numerical
solution with a smaller number of mesh points, i.e. with a faster
algorithm. Suppose that the grid increment $h=\Delta x$ of the
spacial variable $x$ is non constant and denote by $h_j=x_{j+1}-x_j$
the $j$th grid increment, where the $x_j'$s are grid points in the
$x-$direction. The number $q_j=h_j/h_{j-1}$ is often called non
uniformity parameter of the difference grid \cite{andreev}. A direct
substitution of the non constant increment $h_j$ into the FDS
\eqref{eq:3.1} gives rise to the nonuniform version of equation
\eqref{eq:3.3}. A similar but much simpler discretization of
equation \eqref{eq:2.10} can be derived for a nonuniform time
stepping. For the nonuniform discretization of either of the
variables, we used a geometric progression. Thus for the space
variable, we set $h_{1}=h,~h_j=R^{j-1}h$ for some positive numbers
$R$ and $h.$ Here, $R$ turns out to be the non uniformity parameter
and the $h_j$'s form a geometric progression with common ratio $R.$
Values of $R$ not close to $1$ will lead to instability of the
numerical solution, while for $R$ close to $1$ the resulting scheme
will tend to be stable \cite{andreev}.\par

A two-dimensional grid made up of $M$ space intervals and $L$ time
intervals will be referred to as  a grid of mesh (size) $M \times
L.$ In the case of the ${\sc fds}$ that we are considering, $L$ and
$M$ are to be so chosen that \eqref{eq:kh} is satisfied, and this is
easily achieved by letting $L=M.$ In some very rare cases (e.g when
 $\sigma \geq 35\%$), the occurrence of negative option prices at the
intermediary time steps might become persistent and slow down the
convergence of the scheme. This type of difficulty is also easily
resolved by letting $L$ be slightly larger than $M,$ say $L= 1.5
M.$\par

 It should be noted that our algorithm for the implementation of the
optimal boundary condition shares some similarities with the method
for finding the conditional expectation estimator in the Monte Carlo
simulation of a down-and-in barrier option \cite{ross}. Indeed, in
such a Monte Carlo valuation, as soon as the barrier is breached,
the simulation run is ended, and the corresponding estimator is the
Black-Scholes value for the resulting option parameters. In the same
way for the finite difference scheme, as soon as we are certain that
the initial price is sufficiently far away for the barrier to become
worthless, the corresponding option value is its Black-Scholes
value.\par

{
%
}
\subsection{Discretely monitored barriers}\label{s:discrete}
\par
In the case of discretely sampled barriers, we assume that barrier
monitoring occurs either on a daily basis or on a weekly basis. We
adopt a day count convention which will allow a comparison of our
numerical results with others at our disposal by letting a year
consist of $50$ weeks, and a week to consist of $5$ days, so a year
has $250$ days. For the numerical solution, we let $\rho$ be the
number of time steps implemented between two consecutive monitoring
dates, so that the total number of time steps is given by $L=N\rho,$
where $N$ is the number of monitoring dates during the option's
life. As is well-known \cite{boyle98, cheuk}, for increased accuracy
and improved convergence, we shall place the barrier itself midway
between grid points.

\section{Numerical results}
\label{s:nres}
 We start by showing the result of the performance test for the
proposed optimal truncation of the solution domain. In most papers
where approximate conditions of the form \eqref{eq:2.6} are used to
determine boundary conditions in barrier option pricing problems,
the value used for $S_{\rm max}$, the maximum asset price in the
solution domain, is usually not indicated. That is the case for
instance in  \cite{zvan} and  \cite{boyle98}. However, by rewriting
and implementing the same algorithm described in the latter paper
under the scheme termed modified explicit finite difference ({\sc
mefd}) in that paper, we've found by trial and error that choosing
$S_{\rm max} = 2 S_0 + 200$ gives exactly the same results as those
in Table 1 and Table 4 of the said paper. This value of $S_{\rm
max}$ also turns out to give the best results for the parameter sets
considered in that same paper. Other choices for $S_{\rm max}$ which
are commonly used  for a similar range of data are $S_{\rm max} = 2
S_0,$ or $S_{\rm max}= S_0 + 100$, as in \cite{clarke}  and
\cite{mayo}. However, inappropriate choices for $S_{\rm max}$ lead
to very poor results as we show in Table ~\ref{tb:boyle}. \par

\begin{table}[htb]
 \caption{\label{tb:boyle} \protect \footnotesize
Comparison of the accuracy and the CPU time under the optimal
boundary explicit scheme ({\sc obes}) and  the {\sc mefd} of
\cite{boyle98} for a down-and-out call. The fixed parameter set is
$T~=~1, K~=~100, B~=~90,\sigma~=~0.25, r~=~0.10.$ Numbers within
parentheses are CPU times.}
\begin{tabular}{rcrcrcrcrcrcr}
\hline \\[0.3mm]

\parbox[b]{0.3in}{Mesh\\
(L)} &&\multicolumn{5}{c}{\parbox[b]{1.5in}{$S_0=95$\\Closed-Form:
$5.9968$ }} &&\multicolumn{5}{c}{\parbox[b]{1.5in}{$S_0=91
$\\Closed-Form: $1.2738$}}\\ \cline{3-7}
\cline{9-13}\\[0.25mm]

&&{\sc obes}&&\multicolumn{3}{c}{\sc mefd} &&{\sc
obes}&&\multicolumn{3}{c}{\sc mefd}\\\cline{5-7} \cline{11-13} \\[0.0009mm]
&&&&$S_{\max}=$&&$S_{\max}=$
&&&&\multicolumn{3}{c}{$S_{\max}=$}\\
&&&&$2S_{0}+200$&&$2S_{0}$
&&&&\multicolumn{3}{c}{$2S_{0}+200$}\\\hline
                   &&&&&&&&&&&&\\
50  &&6.01109  && 6.0111  && 6.8299 && 1.2654  && 1.2654  && \\
    &&(0.36)   && (0.344) &&        &&(0.39)   && (0.33)  && \\
100 && 5.9984  && 5.9984  && 6.8603 && 1.2704  && 1.2704  && \\
    && (0.53)  && (0.44)  &&        && (0.53)  && (0.47)  && \\
150 && 5.9965  &&5.9997   && 6.7449 && 1.2719  && 1.2719  && \\
    &&(2.09)   &&(2.33)   &&        && (2.203) && (2.42)  && \\
200 && 5.9993  &&5.9993   &&6.6897  && 1.2725  && 1.2726  && \\
    && (2.54)  &&(2.70)   &&        && (2.58)  &&(2.85)   && \\
700 &&5.9970   &&5.9970   && 6.74   && 1.2738  && 1.2738  && \\
    &&(15.81)  &&(17.766) &&        &&(15.75)  && (18.16) && \\
800 &&5.9972   &&5.9972   && 6.78   &&1.2739   &&1.2739   && \\
    && (19.20) &&(21.906) &&        && (19.44) && (22.30) &&\\
900 &&5.9973   &&5.9973   &&6.80    &&1.2739   && 1.2739  && \\
    &&(10.703) &&(12.047) &&        &&(10.922) &&(11.625) && \\
1000&&5.9969   &&5.9969   &&6.7991  &&1.2739   &&1.2739   && \\
    &&(33.86)  &&(40.59)  &&        &&(34.31)  &&(42.25)  && \\
2000&&5.9970   &&5.9970   &&6.7949  &&1.2738   &&1.2738   && \\
    &&(72.67)  &&(85.97)  &&        &&(75.12)  && (85.797)&& \\
3000&&5.9969   &&5.9969   &&6.7718  &&1.2738   &&1.2738   && \\
    &&(189.34) &&(238.141)&&        &&(192.062)&&(236.66) &&\\
4000&&5.9969   &&5.9969   &&6.8020  && 1.2738  &&1.2739   && \\
    &&(263.28) &&(320.187)&&        &&(268.39) &&(335.31) &&\\
    &&         &&         &&        &&         &&         &&\\\hline
\end{tabular}
\end{table}

 Contrary to conditions of the form given by equations
 \eqref{eq:4.2}-\eqref{eq:4.3}
that yield a systematic determination of the optimal boundary
conditions, there is certainly no formula available for the best
choice of $S_{\rm max}$ and it's value is usually determined only by
trial and error.
  For two different values of $S_0,$ Table ~\ref{tb:boyle} compares
  for a down-and-out barrier option
  the convergence rates under the {\sc mefd} scheme  and under
  our optimal boundary explicit scheme ({\sc obes }) using data from  Table
  1 of  \cite{boyle98}. The two different formulas used for
  $S_{\rm max}$ are
  $S_{\rm max}=2 S_0 + 200$ and $S_{\rm max} =2 S_0.$ The parameter set is
  {$ T~=~1, K=100,  B=90,  \sigma =0.25,$ and $r=
  0.10$}. For the explicit scheme used, the mesh size is determined
  by a single parameter $L$ representing the number of required time
  steps, because of the usual stability conditions relating the time and
  the space increments in an explicit scheme. Moreover, this
  constraint for the {\sc mefd} scheme upon which all explicit schemes
  considered in this paper are based is given by
\begin{equation} \label{eq:5.1}
  \Delta y = \lambda  \sigma \sqrt{\Delta t},
\end{equation}
  where $y= \log(S).$
Numbers within parentheses in Table ~\ref{tb:boyle} are CPU times.
All runs in this paper were carried out on a conventional Pentium
Celeron 2.40 Ghz CPU. In all the tables the parameters $q$ and $Rb$
are assumed to be zero, unless otherwise indicated.\par

Columns 2 and 3 of Table ~\ref{tb:boyle} show that the {\sc obes}
and the {\sc mefd} schemes give essentially the same option values
for the given parameter set, when $S_{\rm max}= 2 S_0 + 200$ in the
{\sc mefd}. However, our {\sc obes} appears to be up to $20 \%$
faster
 for all significant number of time steps (i.e. for $L \geq 150$).
 For  small values of $L$ below $150,$ the excess amount of time
 used in the {\sc obes} to calculate the option values at different
 time steps on the upper boundary determined by $S_m$ exceeds the
 excess time required in the {\sc mefd} to implement the additional time
 steps. But this is of no importance as regards the performance of
 the schemes, as for explicit schemes accuracy is generally not
 achieved below $L=1000$ steps. The same conclusion applies when
 comparing columns 5 and 6 of the table.\par
   A comparison of Column 3 and 4 of the same table shows the
importance of choosing an appropriate value for $S_{\rm max}.$
Indeed, the same calculations in Column 3 when done in Column 4 with
$S_{\rm max} = 2 S_0,$ which as already mentioned is a commonly used
value for similar parameter sets in option pricing problems, reveals
that the scheme certainly does not even converge to the closed-form
value of $5.9968.$  However, even for those values of $S_{\rm max}$
for which the {\sc mefd} does converge, there is no systematic way
for finding them. \par

\begin{table}[h]
 \caption{\label{tb:volat} \protect \footnotesize Effect of
volatility on the performance of the {\sc obes} and the {\sc mefd}
of \cite{boyle98} for a down-and-out call with fixed parameters
$S0~=~95, T~=~1, K~=~100, B~=~90,$ and  $r~=~0.10.$}
\begin{tabular}{ccc ccc}
\hline\\[1pt]
  & Mesh & {\sc obes}  & \multicolumn{2}{c}{\sc mefd}  &\parbox[b]{0.4in}{Closed-\\[-4.3mm]form}
  \\\cline{4-5}\\[0.0009mm]
   & & & \parbox[b]{0.9in}{\footnotesize $S_{\max} \\=$ $2 S_0+ 200$}      &   \parbox[b]{0.5in}{\footnotesize   $S_{\max}\\=$ $2 S_0$} & \\\hline
            &&& &&\\
\multirow{2}{*}{$\sigma=0.25$} &200 $\times$ 200  &5.9993 &5.9993
&6.6897  &
\multirow{2}{*}{5.9968}\\[1.5mm]
&2000 $\times$ 2000&5.9970 &5.9970 & 6.7949 &\\[3mm] \hline\\[0.009mm]
     \multirow{2}{*}{$\sigma=0.30$}
     &200 $\times$ 200   &5.9071 &5.9075 &7.6180  &\multirow{2}{*}{5.9060}
     \\[1.5mm]
     &2000 $\times$ 2000 &5.9061 &5.9065 &7.7124  &
     \\[3mm] \hline \\[0.09mm]
          \multirow{2}{*}{$\sigma=0.40$}
          &200 $\times$ 200   &5.7498 &5.7802 &9.2320
          &\multirow{2}{*}{5.7502}\\[1.5mm]
          &2000 $\times$ 2000 &5.7503 &5.7789 &9.5294  &     \\
          &                   &       &       &        &     \\\hline
\end{tabular}
\end{table}

  The validity of our optimal determination of boundary conditions
is further demonstrated by the effect of volatility in Table
\ref{tb:volat}. Indeed, this experiment shows that as the volatility
increases, the {\sc mefd} becomes more and more inaccurate for
almost every possible choice of $S_{\rm max},$ and does not even
tend to converge for values of $\sigma $ above $40 \%.$ On the other
hand the {\sc obes} is perfectly unaffected by any change in the
value of $\sigma.$ Indeed, for large values of $\sigma,$ the
approximation
 $f(S,t) \sim S$ for all $t$ is only tolerated in the {\sc mefd} for
much larger values of $S.$ We've found that by adequately increasing
the value of $S_{\rm max}$ for a larger value of $\sigma,$
convergence under the {\sc mefd} can still be achieved, but at the
expense of much longer computing times and additional trial and
error experiments to find $S_{\rm max}$. Some complications might
also arise in the choice of $S_{\rm max},$ due to the constraint
\eqref{eq:5.1}. As for the {\sc obes}, any change in the value of
$\sigma$ is fully captured by both $S_m$ and the corresponding
Black-Scholes value, so that the {\sc obes} remains totally
unaffected by such changes.\par
\begin{table}[h]
 \caption{ \label{tb:bound} \protect \footnotesize Down-and-out option
values for initial asset prices closest to the upper boundary (U),
the lower boundary (L) or at midway (M). The fixed parameters are
$K~=~150,$ and $B~=~180.$ Numbers within parentheses are CPU times.}
\begin{minipage}[c]{ \textwidth}
\begin{tabular}{c|cc ccc ccc}
\hline
                &&& &&& &&\\
 &&\multicolumn{3}{l}{\parbox[b]{1.2in}{$T=0.25,~\sigma=0.20,\\r=0.05$}}& & &
 \multicolumn{2}{l}{\parbox[b]{1.2in}{$T=1,~\sigma=0.45,\\r=0.07$}}\\\cline{2-5}\cline{7-9}\\[1pt]
 &&\multicolumn{3}{l}{$S_{m}=271.906$}&&&\multicolumn{2}{l}{$S_{m}= 1345.08$}\\
 &&\multicolumn{3}{l}{$x_{m}-x_{b}=0.4125$}&&&\multicolumn{2}{l}{$x_{m}-x_{b}=2.0112$}\\ \cline{2-5}\cline{7-9}\\[1pt]
 &$S_{0}$ &Closed-&{\sc hobis}    &{\sc habis}   & &$S_0$& Closed-  &{\sc hobis}\\
 &        &Form    &Mesh          &Mesh & &   & Form     &Mesh\\\cline{1-5}\cline{7-9}\\[1pt]

\multirow{5}{*}{U}
& 271.905 &123.768 &3 $\times$1 &25 $\times$ 25 &  &1345.07 &   1205.21&2 $\times$ 1\\
&         &         &(0.00)       &(0.05)        &  &&         &             \\
 &271.902  &123.765 &3 $\times$ 1 &25 $\times$ 25 &&1345.00& 1205.14&3 $\times$ 1\\
 &  270.000  &121.862 &10 $\times$ 10 &25 $\times$ 25 &&  1344.00& 1204.14 &7 $\times$ 7\\
 & 265.000  &116.861 &12 $\times$ 12 &25 $\times$ 25 && 1340.00&1200.14&10 $\times $
 10\\ \hline
 &         &         &             &               &&         &            \\
\multirow{5}{*}{L}
 &180.001 &0.0026 &6 $\times$ 6 &30 $\times$ 30 & &180.001&0.0015 &3 $\times$ 1 \\
 &        &       &  (0.03)     &               & &&         &               \\
 &180.010  &0.0264   &11 $\times$ 11 &30 $\times$ 30&&180.01&0.0015 &3 $\times$ 1 \\
 &181.000 &2.6297&135 $\times$ 135&400 $\times$ 400& &181.00&1.4817 & $95\times$ 95 \\
 &       &      & (1.20)         &(14.25)         &&&       &(0.55)
 \\ \hline
 &         &         &            &              &&&          &            \\
\multirow{2}{*}{M}
 &225.953 &77.2335 &95 $\times$ 95 &600$\times$600&&762.54&622.632&$85\times85$\\
 &        &        &(0.53)           &(39.26)       &&&       & (0.52) \\[1mm]\hline
\end{tabular}
\end{minipage}
\end{table}

   The efficacy of combining the high-order accurate implicit scheme
determined by equation \eqref{eq:3.2}, with the stretching
coordinates transformation given by \eqref{eq:2.7}-\eqref{eq:2.8},
and the optimal truncation of the solution domain is first tested
for a down-and-out barrier call option in Table ~\ref{tb:bound}. We
denote by {\sc hobis} (high-order optimal boundary implicit scheme)
the high-order implicit finite difference scheme that applies the
optimal determination of the boundary conditions, while the
corresponding scheme in which a boundary condition is determined by
the approximate condition \eqref{eq:2.6} will be referred to as {\sc
habis} (high-order approximate boundary implicit scheme).\par
   A striking feature of the high-order accurate {\sc hobis} is
its fastest convergence for initial asset prices closest to
boundaries of the solution domain. For initial asset values closest
to $S_m$ on the upper boundary (U), the minimum mesh size that
returns the anlytical option price is  $2 \times 1$ and corresponds
to a CPU time of $0.00$ seconds. For values closest to the barrier
on the lower boundary (L), the minimum mesh size is $7 \times 7$ and
corresponds to a CPU time of $0.02$ seconds. This performance is
clearly due to the exactness of option values at the boundaries in
combination with the fact that interpolation yields best results for
points closest to known function values. But this is also a
consequence of the coordinates transformation (in the first set of
data with $\sigma = 0.20$ in Table ~\ref{tb:bound}, $x_m- x_b=0.4125
<1$), the higher-order accuracy of the scheme that forces all error
less than one to quickly disappear, and the implicit nature of the
scheme that also tends to quickly reduce any error. For instance, in
the second set of data in Table \ref{tb:bound} with $\sigma= 0.45,$
and $S_0= 1345.00,$ although the space increment is
$$\Delta x= 2.011/2 =1.005 > 1, $$
all significant errors have disappeared by the completion of the
first time step. The table also shows as expected that the more we
move away from the boundaries, the higher the error tends to become
in option values with a fixed mesh size. We have therefore also
indicated the minimum CPU time required for initial values midway
(M) between the boundaries, and we can reasonably argue that for any
given scheme and any parameter set, the maximum CPU time for all
values of $S_0$ in the solution domain is close to the maximum CPU
time obtained for $S_0$ close to the boundaries and at midway. This
means for instance that for the first set of data with $\sigma
=0.20,$ and the second set of data with $\sigma = 0.45,$ to within
four significant digits of accuracy, the maximum CPU time is close
to $1.20 s$ and and $0.55 s,$ respectively. It should be noted that
this is also the maximum CPU time for any possible value of the
initial price. Indeed, another feature of the optimal boundary
condition is to determine exactly when a barrier becomes worthless
for a given value of $S_0,$ so its value is reduced to the
corresponding vanilla Black-Scholes value. More precisely, by
construction of $S_m,$ whenever $S_0 \geq S_m$ the barrier option
value is the vanilla Black-Scholes value, and so any scheme
(explicit, implicit or otherwise) based on the optimal boundary
returns the option value in zero seconds for such values of $S_0.$
\par
%
\begin{figure}[ht]
\begin{center}
\caption{\label{fg:dbar1} \protect \footnotesize Convergence
patterns under the {\sc hobis} for a double knock-out. The parameter
set is $S_0~=~100, T~=~0.5, K~=~100, B_l~=~75, B_u~=~125,
\sigma~=~0.20$ and $r~=~0.10.$ All absolute errors are less than $10
\%$ with a $17\times 17$ mesh, and less than $2\times10^{-3}$ with a
$ 70 \times 70$ mesh. \vspace{2mm}}
\includegraphics[width= \textwidth]{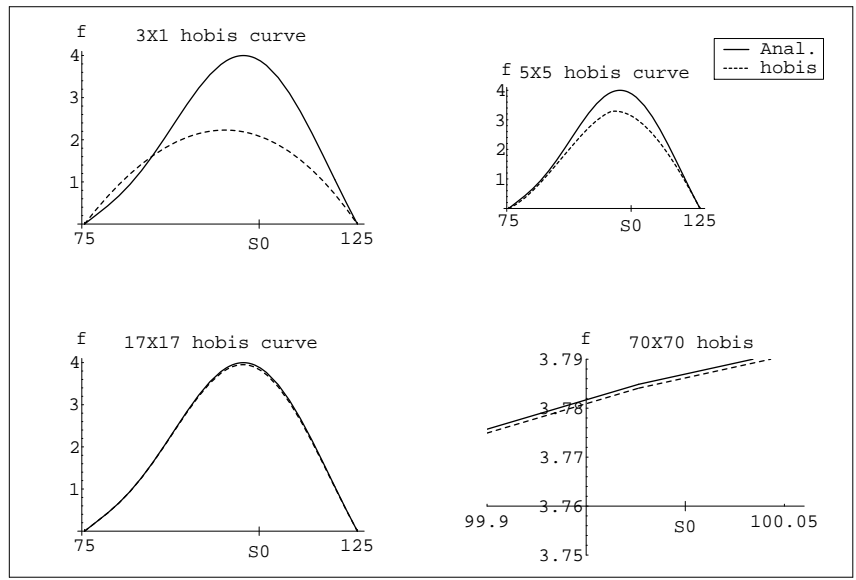}
\end{center}
\end{figure}

The $5$th column of Table ~\ref{tb:bound} shows that implementing
the same high-order accurate implicit scheme by using the
approximate boundary condition of the form \eqref{eq:2.6} leads to
much more inaccurate results and requires about $10$ times more
computing time, although the value of $S_{\rm max} = 2 S_0 + 200$
appears to be the best choice for all sets of parameters considered.
This numerical experimentation confirms the fact that the superior
accuracy of the {\sc hobis} is valid for all finite difference
schemes.\par
   The {\sc hobis} curves depicted in Figure \ref{fg:dbar1} for a double
knock-out also display a similar trend of convergence described in
Table \ref{tb:bound}, in which accuracy is much quickly  achieved
for points closer to the boundaries of the solution domain, while
the middle point returns the most coarse value (at least for the
first few mesh sizes). The parameter set for this figure is
$S_0=100, T=0.5, K=100, B_l~=~75, B_u~=~125, \sigma~=~0.20$ and
$r=0.10.$  The figure clearly shows the global accuracy of the {\sc
hobis} for different mesh sizes. For a $17 \times 17$ mesh size, the
{\sc hobis} curve is essentially indistinguishable from the
analytical curve and the maximum absolute error is  $5.1\times
10^{-2}.$  This error reduces to $3.0 \times 10^{-3}$ for a $70
\times 70$ mesh size.\par
%
\begin{table}[h]
  \caption{ \label{tb:maxer} \protect \footnotesize Maximum absolute error under the
  {\sc hobis} for continuously monitored down-and-out and double
  knock-out calls. The fixed parameter sets are $ K=100, B=90,
  r=0.10$ for the down-and-out call and $K=100, Bl=75, Bu= 125,
  r=0.10$ for the double knock-out call. The maximum absolute error
  is that on the $M+1$ computed option prices for any given mesh
  size of the form $M \times L.$}

\begin{minipage}[l]{\textwidth}
\begin{tabular}{l l c c c l c c}\hline \\[0.15mm]
\multicolumn{1}{c}{\parbox{0.7in}{}}& \multicolumn{3}{c}{ \sc
down-and-out call} & &\multicolumn{3}{c}{\sc double knock-out call}
\\[1mm] \cline{2-4} \cline{6-8}\\[1mm]

$\sigma$  & \parbox{0.5in}{para-\\meters}  & { mesh}  & { error}  & & \parbox{0.5in}{para-\\meters}  & { mesh}&{ error}\\[2mm] \hline \\[0.15mm]
\multirow{5}{*}{$0.15$}
& $T=0.5$    & $7 \times 7 $   &$2.4 \times 10^{-2}$ & &$T=0.5$ &$20 \times 20$  &$7.9 \times 10^{-2}$  \\[1pt]
& $T=0.5$    &$20 \times 20$   &$1.3 \times 10^{-3}$ & &$T=0.5$ &$100 \times 100$&$3.1 \times 10^{-3}$  \\[1pt]
& $T=0.5$    &$100 \times 100$ &$3.0 \times 10^{-5}$ & &$T=0.5$ &$200 \times 200$&$7.8 \times 10^{-4}$  \\[1pt]
& $T=1$      &$100 \times 100$ &$3.6 \times 10^{-5}$ & &$T=1$   &$200 \times 200$&$4.0 \times 10^{-4}$  \\[5pt]
&\parbox[c]
 {0.5in}
 {$\{T=1,$\\
$ Rb=3\}$} & $100 \times 100$ &$5.6 \times 10^{-5}$ &
& $T=0.1$&$200\times200$   & $3.5 \times 10^{-3}$\\[1mm] \hline \\[2pt]
$0.35 $& $T=1$    & $100 \times 100$&$2.6 \times 10^{-4}$ &   &$T=1$
&$200 \times 200$ & $6.3 \times 10^{-2}$ \\\hline
\end{tabular}
\end{minipage}
\end{table}
%
For almost all typical model parameters, the {\sc hobis} and
analytical curves are essentially indistinguishable in the case of a
down-and-out call, even for the smallest $3 \times 1$ mesh size.
We've therefore given in Table \ref{tb:maxer} the maximum absolute
error on computed option prices for some given mesh sizes, and for
both a down-and-out call and a double knock-out call. The table
shows that the absolute {\sc hobis} error is strictly less than
$10^{-2}$ for a $20 \times 20$ mesh when the option is a
down-and-out, and that global accuracy is much quickly achieved with
a down-and-out than with a double knock-out. On the other hand there
does not seem to be a simple correlation between accuracy and expiry
date for a fixed mesh size in the case of the double knock-out call,
at least for the three different expiry dates considered. However,
it clearly transpire from the table that the {\sc hobis} is very
sensitive to volatility, especially in the case of a double
knock-out call where the maximum absolute error is about $100$ times
larger when the volatility goes from $0.15$ to $0.35.$ The table
also shows that accuracy is a bit more expensive when the
computation of option values incorporates a rebate payment.\par
%
\begin{figure}[ht]
\begin{center}
\caption{\label{fg:cp3schem} \protect \footnotesize Error in the
computed option prices $f$ under the {\sc hobis}, the
Crank-Nicolson, and the fully implicit schemes. The parameter set is
$T=0.5, K=100, B=90, \sigma =0.20$ and $r=0.10.$ The mesh size is
$40 \times 40$.}
\includegraphics[width=0.70\textwidth]{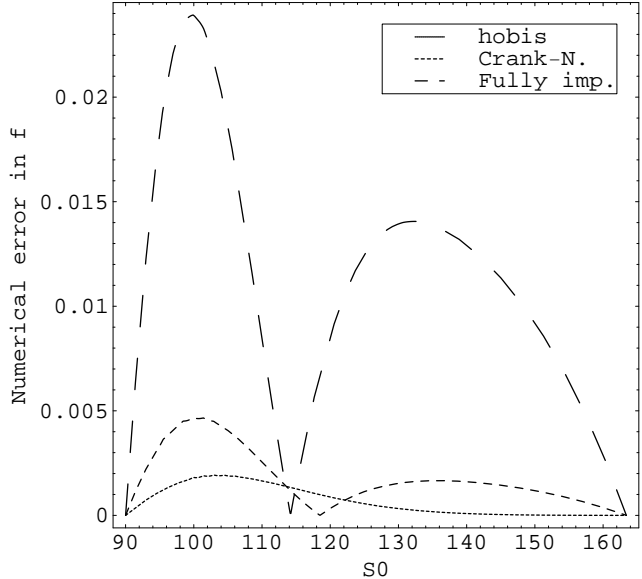}
\end{center}
\end{figure}
%
%
%
%
\begin{figure}[h]
\begin{center}
\caption{\label{fg:cp2schem} \protect \footnotesize Error in the
computed option prices $f$ under the {\sc hobis}, and the
Crank-Nicolson schemes. The parameter set is $T=0.5, K=100, B=90,
\sigma =0.20$ and $r=0.10.$ The mesh size is $40 \times 40$.  .}
\includegraphics[width=0.60\textwidth]{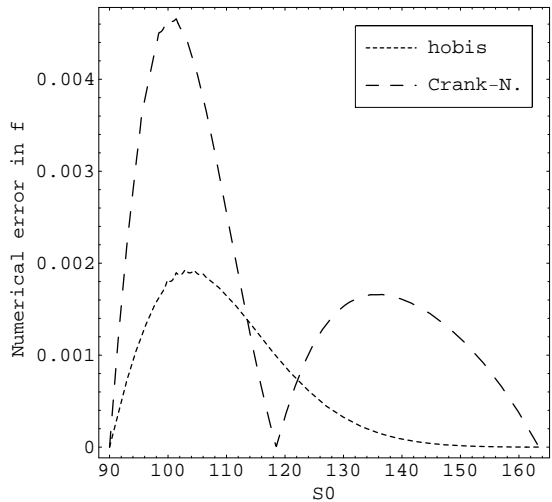}
\end{center}
\end{figure}
Although the {\sc hobis} is fourth-order accurate in the space
variable and second-order accurate in the time variable while the
popular Crank-Nicolson scheme is second order accurate in both the
space and the time variables, this does not necessarily precludes
the actual numerical implementation of these schemes to prove
otherwise. We've therefore plotted against the initial price $S_0$
in Figure \ref{fg:cp3schem}  the numerical errors in the computed
down-and-out option prices under the {\sc hobis}, the
Crank-Nicolson, and the Fully implicit schemes. The latter scheme
which was used for option pricing in \cite{zvan} is only first-order
accurate in time. The figure shows that the {\sc hobis} is much more
accurate than the Crank-Nicolson scheme for the corresponding mesh
used, and the fully implicit scheme performs very poorly compared to
the {\sc hobis}. In fact the maximum absolute errors resulting from
these plots are $0.00193$ for the {\sc hobis}, $0.00466$ for the
Crank-Nicolson, and $0.02392$ for the fully implicit scheme. The
error functions for the last two schemes have each a singularity
where the error tends to jump to zero, but this has essentially no
effect on their performance. We have however, plotted the same error
functions in Figure \ref{fg:cp2schem} for the two best performing
schemes, namely the {\sc hobis} and the Crank-Nicolson scheme, to
clarify the difference between their levels of accuracy. Although
the only mesh size used for these two figures is $40 \times 40,$ our
experimentations shows that this trend persists for all mesh sizes
larger than say, $20 \times 20.$ The common parameter set for the
two figures is $T=0.5, K=100, B=90, \sigma=0.20$ and $r=0.10.$ \par
  Table  \ref{tb:zvan} compares {\sc hobis} values with those
  reported in Table 1 and
Table 2 of \cite{zvan} for a down-and-out and a double knock-out
call, and corresponding to values  obtained in \cite{zvan} using a
fully implicit method of first order of accuracy ({\sc zvan}), and
in \cite{cheuk} using a trinomial tree method ({\sc c-v}). The
options parameters used are $S=100,\, T=0.5,\, K=100,\, \sigma=
0.20,\, r=0.10.$ For the down-and-out, $B= 99.9,\,$ while for the
double knock-out, the lower barrier is set at $B_l= 95$ and the
upper barrier is set at $B_u= 125.$ For continuously monitored
barriers, the {\sc hobis} values correspond to the analytical values
to within the required accuracy, as are those computed by Cheuk and
Vorst, while surprisingly the {\sc zvan} values of \cite{zvan} do
not coincide with analytical values. Closed-form solutions for
continuously monitored barrier options have been derived by various
authors \cite{merton, rub, kuni, geman, pelsser}. The {\sc hobis }
appears not only to be more acurate, but also about twice faster
than the implicit scheme of \cite{zvan}. In the case of discrete
monitoring, the {\sc hobis} values are essentially the same as the
{\sc c-v} values to within the required accuracy, but differ
significantly again from those obtained using the fully implicit
method of \cite{zvan}. In general accuracy is very costly in terms
of CPU time for discretely monitored barrier options and the
computing time in Table ~\ref{tb:zvan} for discrete monitoring
corresponding to the {\sc zvan} column looks too little. For similar
amounts of time the {\sc hobis} returns similar option values, but
does not converge to those values, and it generally requires up to a
hundred of seconds to converge significantly. \par
\begin{table}[h]
  \caption{ \label{tb:zvan} \protect \footnotesize Down-and-out and
double knock-out call values with continuous and discretely applied
constant barriers.}

\begin{minipage}[l]{\textwidth}
\begin{tabular}{r c c r c c c r}\hline \\[0.15mm]
\multicolumn{1}{c}{\parbox{0.7in}{Monitoring \\frequency}}&
\multicolumn{3}{c}{ Down-and-out Call} & &\multicolumn{3}{c}{Double
Knock-out Call}
\\[1mm] \cline{2-4} \cline{6-8}\\[1mm]

  & {\sc hobis}  & {\sc zvan}  &{\sc c-v}  & &  {\sc hobis}  & {\sc zvan}&{\sc c-v}\\[2mm] \hline \\[0.15mm]
Continuous &0.165  &0.164  & 0.165 & &2.033   &2.037  & 2.033 \\
      &(0.03) &(0.05) &       & &(0.240) &(0.48) &  \\[3mm]
Daily &1.511  &1.506  & 1.512 & & 2.482  &2.485  &  2.482\\
      &     &(13.96)&       & &     &(37.93)& \\[3mm]
Weekly &3.008 & 2.997 & 2.963 & & 3.006     &3.012  & 2.989\\
       &    & (2.80)&       & &      &(9.47) & \\\hline \\[1mm]
\end{tabular}
\end{minipage}
\parbox[t]{0.9\textwidth}{  \protect \footnotesize  The fixed options parameters are
$S~=~100,\, T~=~0.5,\, K~=~100,\, \sigma~=~ 0.20,\\ r~=~0.10.$ For
the down-and-out, $B= 99.9,\,$ while for the double knock-out,
$B_l~=~ 95, \; B_u= 125.$ {\sc c-v} denotes results obtained in
~\cite{cheuk} while Zvan denotes results obtained results obtained
in \cite{zvan}. Numbers within parentheses are CPU times}
\end{table}

   Although Table ~\ref{tb:zvan} clearly demonstrates the higher
performance of the {\sc hobis} in terms of convergence and accuracy
for continuously monitored options, it is hard to make the same
conclusion from this table about discretely applied barriers, and
this is largely due to the non availability of exact solutions in
such cases. \par
   We have therefore compared in Table ~\ref{tb:disct} the
{\sc hobis} values for a discretely monitored down-and-out option
with those taken from Table 2 of \cite{fusai}. Fusai {\em et al}
\cite{fusai} have reduced the valuation problem for discretely
monitored down-and-out barrier options to a Wiener-Hopf integral
equation, and given a formal  inverse $z$-transform solution to this
equation in terms of a special function plus infinite sums of simple
functions. This however does not give rise to exact option values,
as the solutions need to be evaluated numerically. In addition to
the Wiener-Hopf ({\sc wh}) method of \cite{fusai} the other
numerical methods which are compared with the {\sc hobis} method in
Table  \ref{tb:disct} are the Markov Chain method ({\sc MCh}) of
\cite{MCh}, the trinomial tree ({\sc tt}) of \cite{TT}, the Simpson
recursive quadrature method ({\sc sq}) of \cite{SQ}, and the Monte
Carlo simulation method ({\sc mc}) of \cite{MC} with antithetic
variables  and $10^8$ simulation runs. Table ~\ref{tb:disct} shows
that the {\sc hobis} values are much closer to the {\sc wh} values
of \cite{fusai} which is the only numerical method derived from an
analytical solution. The table also shows that apart from the {\sc
tt} values in column $6,$ there is in general a good level of
agreement between all the numerical methods. It should be noted that
values from the {\sc tt} method that seems to perform relatively
poorly in the table are exactly the same values reported for the
same discretely monitored barrier option in \cite{cheuk}, some of
which values appear in Table \ref{tb:zvan} above. This shows that
for discrete monitoring, the discrepancies between the {\sc hobis }
values and other values listed in Table ~\ref{tb:zvan} cannot be
perceived as a lack of performance of the {\sc hobis}. In fact Table
~\ref{tb:disct} simply confirms a wide spread perception that tree
methods are less efficient compared to the most common numerical
methods for option pricing. The fixed option parameters in Table
~\ref{tb:disct} are $S~=~100,\, T~=~0.5,\, K~=~100,\, \sigma~=~
0.20,\, r~=~0.10,$ and $N$ and $B$ denote as usual the number of
monitoring dates and the barrier value, respectively.\par
\begin{table}[h]
 \caption{ \label{tb:disct}  \protect \footnotesize  Comparison of the
HOBIS results for a discretely monitored down-and-out call. The
parameter set is $S=100,\, T=0.5,\, K=100,\, \sigma= 0.20,\,
r=0.10,$ and $N$ and $B$ denote as usual the number of monitoring
dates and the barrier value, respectively. }

\begin{minipage}[c]{\textwidth}
\begin{tabular}{r l  l l l l l l}\hline \\[0.15mm]
$N$  & $B$    & {\sc hobis}& {\sc  wh-ir} &{\sc MCh}& {\sc tt} &{\sc
sq} &{\sc mc}
\\ \hline \\[1.0 pt]
25 & 95  &6.63176&6.63156&6.6307&6.6181&6.6317  &6.63204 \\
25 & 99.5&3.35542&3.35558&3.3552&3.3122&3.3564  &3.35584 \\
25 & 99.9&3.00848&3.00887&3.0095&2.9626&3.0098  &3.00918 \\[1.0pt] \hline \\[1.0pt]

125 & 95  &6.16797&6.16864&6.1678&6.1692&6.1687 &6.16879 \\
125 & 99.5&1.96143 &1.96130&1.9617&1.9624&1.9628 &1.96142 \\
125 & 99.9&1.51098 &1.51068&1.5138&1.5116&1.5123 &1.5105 \\ \hline

\end{tabular}
\end{minipage}
\end{table}
Analytical pricing formulas for double barrier options with rebate
payment has recently been obtained in \cite{pelsser} based on the
inversion of the Laplace transform of the probability density
functions by contour integration. Sidenius obtained a similar
solution in \cite{siden} using an approach based on path counting.
These two results are certainly among the first ones incorporating a
rebate payment for continuous barrier options, and actual barrier
option values computed with these formulas hardly exists, partly
because any computation using these formulas is still quite computer
intensive. The scarcity of numerical option values in the scientific
literature for discretely applied barrier options is essentially due
to the lack of exact solutions for such options. We have thus listed
in Table ~\ref{tb:qRb} some numerical values for a number of
knock-out option values that incorporate  dividend ($q$) and rebate
($Rb$) payments. The fixed options parameters are $S=100,\, T=0.5,\,
K=100,\, \sigma= 0.20,\, r=0.10.$ The various knock-out option types
considered are the down-and-out, the up-and-out, and the double
knock-out barrier options. The barrier application is either
continuous or discrete in time. Numbers within parentheses in the
table are CPU times.\par
   In Table ~\ref{tb:qRb} the values of $q$ and $Rb$ are generally
chosen to display some of the effects of these parameters. For
example if we denote by $Cbar$ and $Dbar$ the value of a given
barrier option under continuous monitoring and discrete monitoring
respectively, and by $Van$ the value of the corresponding vanilla
option with same parameters, in the absence of rebates we must have
\begin{equation} \label{eq:barineq} Cbar \leq Dbar \leq Van \end{equation}
and the generally very slow convergence of a discretely monitored
barrier option can be verified using the inequalities in
\eqref{eq:barineq}, given that the absolute error in the computation
of $Dbar$ is at most the difference  $(Van-Cbar)$, and this yields a
better approximation of $Dbar,$ the more $(Van -Cbar)$ is small.
Equation \eqref{eq:barineq} is no longer true when the option pays a
rebate, and many such examples appear in the table. The table also
confirms that no matter what the options' monitoring frequency,
dividend payments tend to drag down the call option value.\par
\begin{table}[h]
 \caption{ \label{tb:qRb} \protect \footnotesize Various {\sc hobis}
option values for continuously monitored and discretely monitored
knock-out calls that incorporate rebate and dividend payments. The
fixed parameters are $S= 100, T=0.5, K=100, \sigma = 0.20,
  r=0.10.$ For discrete monitoring the number of monitoring periods is
  25 and the mesh size used is $4000 \times 6000$.
Numbers within parentheses are CPU times.}
\begin{tabular}{l l l l c c}
\hline \\[1.0pt]

$B_l$ & $B_u$  & q    & $Rb$ & { Continuous} & { Discrete}   \\ \hline \\[1.0pt]

\multicolumn{6}{c}{Down - and - Out} \\[1.5mm]
90      & -     & 0.05 & 3    &  7.491 (0.42)  & 7.493\\
   90   & -     & 0.05& 1.125& 6.719 (0.55)    & 6.841\\
   99.9 &-      & 0.05& 3      & 3.106 (0.04)  & 4.924\\
   99.9 &-      & 0   &0       & 0.165 (0.03)  & 3.009\\[2.5mm]

\multicolumn{6}{c}{Up - and -Out}                           \\[2.5mm]
-     &110  &0.20 &0   & 0.225 (1.18)     & 0.344\\
   -  &110  &0.02 &0   & 0.299  (1.15)    & 0.470\\
    - &100.1  &0.0  &0.01& 0.009 (0.02)     & 0.009\\
     -&100.1&0.0  &3   & 2.987 (0.04)     & 2.735\\[2.5mm]

\multicolumn{6}{c}{Double Knock - Out}               \\[2.5mm]
       95&125 &0      &0    &2.033  (0.240)      & 3.008\\
       95&125 &0.04   &6.66 & 7.057  (0.30)      & 7.256\\
       75&185 &0.045  &0    & 6.863  (0.30)      & 6.864\\
       80&120 &0.04   &0    &2.196  (0.4)       & 2.654\\\hline

\end{tabular}
\end{table}
   As far as the computing times are concerned, Table ~\ref{tb:qRb}
shows that the CPU times in seconds for all the three types of
knock-out options considered and for the given parameters are almost
all between $0$ and $0.5$ seconds under continuous monitoring and do
not exceed $1.18$ seconds in any case. The computing times under
continuous monitoring are smaller for values very close to the
barrier, as already observed in Table ~\ref{tb:bound}. As for
discretely monitored options, the computing time to achieve a very
high accuracy can reach five thousand seconds, partly because it
would be time consuming to try to find out the minimum amount of CPU
time required in such case of very slow convergence to achieve a
given level of accuracy.  We have however noted that the computation
of the discretely monitored double knock-out option with $q=0.045$
in the table gives the indicated value of $6.864$ in only $4.20$
seconds which turns out to be the limiting value, and this is truly
a record CPU time for the computation of a discretely applied
barrier option.\par
   All of our nonuniform grid implementations as described in Section
\ref{s:nsol} did not yield any faster convergence or any better
accuracy in the transformed $(x, \tau)$~-~coordinates . As in
Section \ref{s:nsol}, denote by $x_j$ the $j$th grid point in the
$x$~-~direction, and by $h_j= x_{j+1}- x_j$ the $j$th grid
increment, and similarly let $S_j = K e^{x_j}$ and $h^*_j$ be the
corresponding grid point and grid increment in the original
$S$-coordinates. Then we readily see that
$$
h^*_j= (e^{h_j}-1)\cdot S_j,
$$
showing that a uniform grid in the $x$-coordinates (corresponding
to\\ $h_j~=~h~=~ {\rm constant}$) always yields a nonuniform grid in
the original $S$-coordinates. Moreover, such a nonuniform grid is
typically sufficiently refined, since $h$ is assumed to be small and
thus $e^h-1$ is close to zero. For instance, if $h~=~1/ 10^4$ and
$S_j~=~50,$ then $h^*_j~=~ 0.00500025.$ This shows that the
implementation of effective nonuniform grids is not straightforward
under the combination of the coordinates transformation and the
high-order accurate  FDS. For all of our attempts of nonuniform
grids in the $x$- or $\tau$-coordinates, we achieved the best
results only when the non uniformity parameter $q_j ~=~ h_j/
h_{j-1}$ was equal to $1,$ which corresponds to the uniform grids
discussed in this section.


\section{Conclusions}
\label{s:conclu}For the valuation of various  barrier options, we
have used in this paper a finite difference scheme which is
fourth-order accurate in the space variable and second-order
accurate in the time variable. This scheme has been shown to be much
faster and accurate then all of the most common schemes, namely the
Crank-Nicolson, the explicit, and the fully implicit schemes. We've
also given an optimal determination of the boundary conditions,
which is particularly relevant for single-barrier options, but also
for discretely monitored double knock-out options. In this way, with
a reasonable accuracy we have been able to value continuously
monitored barrier options in a fraction of seconds and with a mere
$20 \times 20$ mesh.\par
 The efficiency of the combined
high-order scheme and optimal determination of boundary conditions
was again demonstrated in the difficult problem of valuing
discretely monitored barrier options. Indeed, a comparison of the
 option values obtained using the resulting
scheme with those obtained using five other numerical methods shows
not only a good agreement, but also that our values are much closer
to the sole method based on an analytical solution for discretely
sampled barrier options. Our optimal determination of  boundary
conditions naturally has several applications, and could be used
amongst others  to classify double barrier options into simpler
types of options.
%




\end{document}